\documentclass[
superscriptaddress,
nofootinbib,
 aps,
reprint
]{revtex4-1}

\usepackage{graphicx}
\usepackage{dcolumn}
\usepackage{bm}
\usepackage{amsmath}
\usepackage{color}
\usepackage{braket}
\usepackage{float}
\usepackage{array,url,kantlipsum}
\usepackage{verbatim}
\usepackage{soul} 
\usepackage{color,xcolor}
\usepackage{mathtools, cuted}
\usepackage[mathlines]{lineno}

\newcolumntype{P}[1]{>{\centering\arraybackslash}p{#1}}

\begin{document}
\preprint{APS/123-QED}

\title{Optically Detected Magnetic Resonance of Nitrogen-Vacancy Centers in Diamond under Weak Laser Excitation}

\author{Yong-Hong Yu}
\author{Rui-Zhi Zhang}
\author{Yue Xu}
\author{Xiu-Qi Chen}
\affiliation{Beijing National Laboratory for Condensed Matter Physics, Institute of Physics, Chinese Academy of Sciences, Beijing 100190, China}
\affiliation{School of Physical Sciences, University of Chinese Academy of Sciences, Beijing 100049, China}

\author{Huijie Zheng}
\email{hjzheng@iphy.ac.cn}
\affiliation{Beijing National Laboratory for Condensed Matter Physics, Institute of Physics, Chinese Academy of Sciences, Beijing 100190, China}
\affiliation{Songshan Lake Materials Laboratory, Dongguan, Guangdong 523808, China}

\author{Quan Li}
\author{Ren-Bao Liu}
\affiliation{Department of Physics, Centre for Quantum Coherence, and The Hong Kong Institute of Quantum
Information Science and Technology, The Chinese University of Hong Kong, New Territories, Hong Kong, China}

\author{Xin-Yu Pan}
\affiliation{Beijing National Laboratory for Condensed Matter Physics, Institute of Physics, Chinese Academy of Sciences, Beijing 100190, China}
\affiliation{Songshan Lake Materials Laboratory, Dongguan, Guangdong 523808, China}
\affiliation{CAS Center of Excellence in Topological Quantum Computation, Beijing 100190, China}

\author{Dmitry Budker}
\affiliation{Johannes Gutenberg-Universit{\"a}t Mainz, 55128 Mainz, Germany}
\affiliation{Helmholtz-Institut, GSI Helmholtzzentrum f{\"u}r Schwerionenforschung, 55128 Mainz, Germany}
\affiliation{Department of Physics, University of California, Berkeley, California 94720, USA}

\author{Gang-Qin Liu}
\email{gqliu@iphy.ac.cn}
\affiliation{Beijing National Laboratory for Condensed Matter Physics, Institute of Physics, Chinese Academy of Sciences, Beijing 100190, China}
\affiliation{Songshan Lake Materials Laboratory, Dongguan, Guangdong 523808, China}
\affiliation{CAS Center of Excellence in Topological Quantum Computation, Beijing 100190, China}

\date{\today}

\begin{abstract}
As promising quantum sensors, nitrogen-vacancy (NV) centers in diamond have been widely used in frontier studies in condensed matter physics, material science, and life sciences. In practical applications, weak laser excitation is favorable as it reduces the side effects of laser irradiation, such as phototoxicity and heating. 
Here we report a combined theoretical and experimental study of (near) zero-field optically detected magnetic resonance (ODMR) of NV-center ensembles under weak 532-nm laser excitation. 
In this region, both the width and splitting of the ODMR spectra decrease with increasing laser power. This power dependence is reproduced with a model that accounts for the laser-induced charge neutralization of NV$^-$- N$^+$ pairs, which alters the local electric field environment.
These results are important for the understanding and development of NV-based quantum sensing in light-sensitive applications. 

\end{abstract}

\pacs{Valid PACS appear here}
\maketitle

Spin defects in solids with an optical interface are among the leading platforms for quantum information science and quantum sensing.
Among the many spin defects, the negatively charged nitrogen-vacancy (NV$^-$) center in diamond stands out, exhibiting excellent spin coherence, bright photoluminescence, and large contrast of magnetic resonance signal \cite{gruber1997scanning}. For quantum sensing applications, these features bring remarkable sensitivity in detecting magnetic field \cite{balasubramanian2008nanoscale, maze2008nanoscale, taylor2008high}, temperature \cite{TempBudker, Sun2011APL}, pressure and other physical quantities \cite{doherty2014Pressure, SingleNVElectricFieldSensing}. At the same time, the atomic structure of an NV center enables spatial resolution down to the nanoscale, and the ultra-stable diamond structure is compatible with complex biochemical environments \cite{Degen2014Review}, even under extreme conditions such as megabar pressures \cite{MBarODMR} or sub-Kelvin temperatures \cite{mK_temperature}.
 
Laser excitation provides an efficient and convenient way to establish the polarization and readout of NV electron spin, and has been widely used in NV-based quantum techniques. However, in practical applications, laser excitation also brings undesirable effects.
For example, phototoxicity is an important issue when studying living cells with fluorescent probes. Recent experiments have shown that even weak laser excitation could bring adverse effects to living cells \cite{phototoxicity2017NatMeth, 6D_tracking}.
Another side effect is light-induced sample heating, which is a serious obstacle for temperature-sensitive applications, such as the characterization of superconducting films or magnetic films with low transition temperatures \cite{PRX2014NV_SC, Laser_Heating_SC, mK_temperature,acosta2019color, schlussel2018wide}, and temperature-controlled cell division during early embryogenesis\,\cite{choi2020PNAS}. 

A straightforward strategy to reduce laser-induced side effects is to use laser beams of sufficiently low intensity,
but the intrinsic properties of NV centers, including charge, spin, and optical properties, have been poorly studied under weak excitation.
Jensen \emph{et al.} reported light-narrowing of the microwave power-broadened magnetic resonance signals of NV centers under a small magnetic field, which was attributed to the light-induced reduction of the NV spin relaxation time ($T_1$) \cite{Budker_narrowing}. 
Under optical pumping, the charge conversion of an NV center (between NV$^-$ and NV$^0$) is inevitable \cite{aslam2013photo}, and the randomly distributed charges, either from NV centers or from nearby charge traps, creating internal electric fields around the NV center\,\cite{mittiga2018imaging}. 
Therefore, it is necessary to consider both the charge and spin degrees of freedom together and revisit the magnetic resonance of NV centers under weak laser excitation. 
Recently, Manson \emph{et al} made an important step in this direction and showed that the charge conversion of NV$^-$- N$^+$ pairs is crucial for understanding the optical properties of NV centers in diamond \cite{manson2018nv}.   

In this paper, we report a combined theoretical and experimental study on optically detected magnetic resonance (ODMR) of NV centers under weak green laser (532\, nm) excitation. Ensembles of NV centers in nanodiamonds, microdiamonds, and bulk diamonds are studied. When the excitation laser power is much smaller (one or more orders of magnitude) than the optical saturation power, both the width and the splitting of ODMR spectra decrease as the optical intensity increases. This phenomenon is well interpreted with a model that takes into account the laser-induced charge neutralization of NV$^-$- N$^+$ pairs in diamond.

\begin{figure}
	\includegraphics[width=0.48\textwidth]{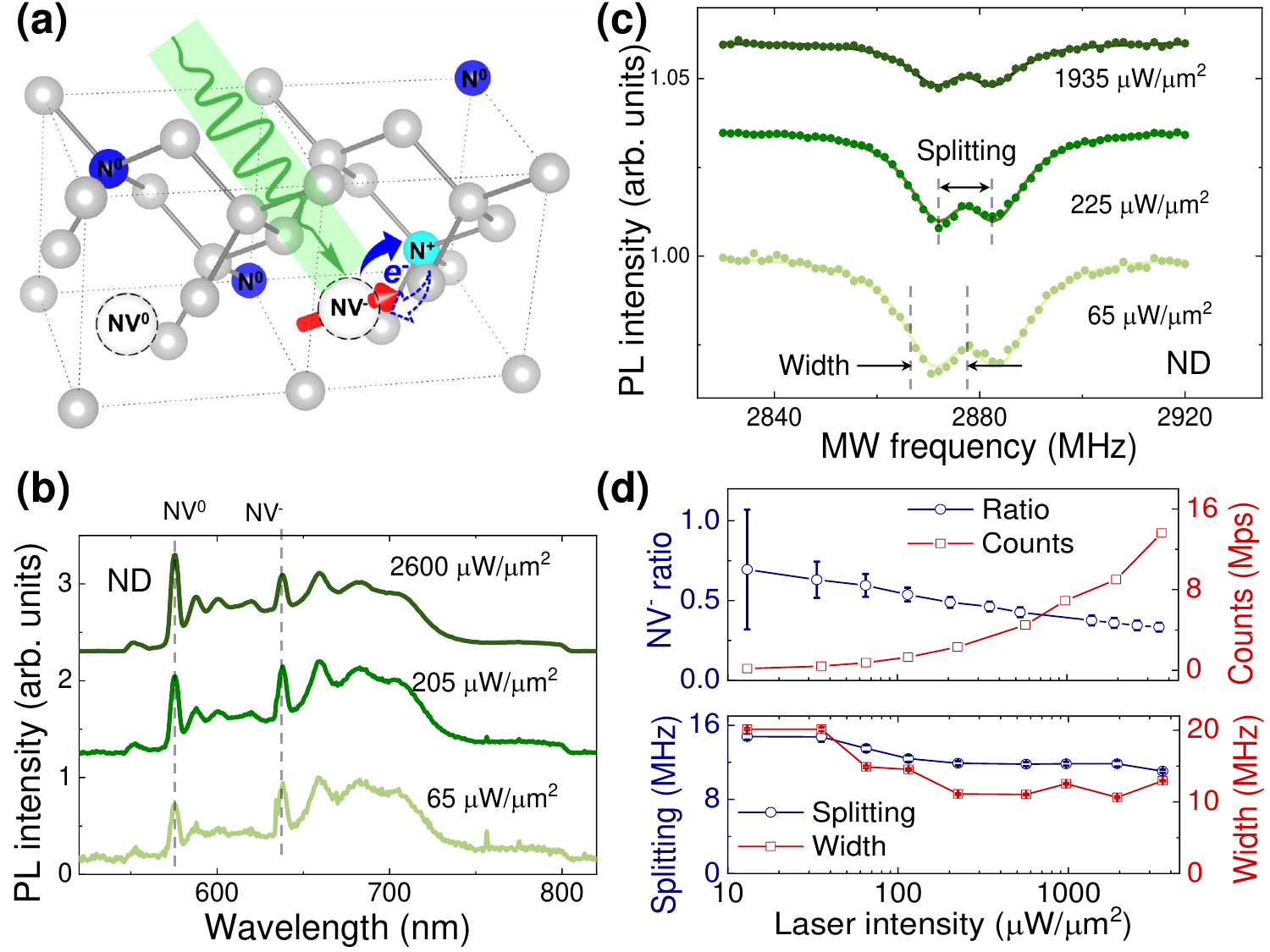}%
	\caption{ \label{fig_1} Laser-induced charge neutralization of NV$^-$- N$^+$ pairs in diamond. (a) Schematic representation of the NV$^-$- N$^+$ model. After the absorption of photons, NV$^-$- N$^+$ pairs could change into NV$^0$- N$^0$ pairs, which leads to a more dilute charge distribution in the diamond lattice and thus smaller internal electric fields acting on the remaining NV centers. (b) Photoluminescence (PL) spectra of a nanodiamond measured at 8.4 K. The curves are vertically offset for better visibility. (c) Zero-field optically detected magnetic resonance (ODMR) spectra of the same nanodiamond, measured at different laser powers. (d) Laser power dependence of the  NV$^-$ ratio and photon count (upper pane), the splitting and width of the ODMR spectra (lower pane).}
\end{figure}

(\emph{Results})
A nitrogen-vacancy center is formed by a substitutional nitrogen atom and a nearby vacancy in the diamond lattice. It has two common charge states, NV$^-$ and NV$^0$, with featured zero-phonon lines (ZPL) at 637\, nm and 574\, nm, respectively.
The negatively charged NV$^-$ center is a natural spin qubit, it exhibits spin-dependent optical transitions that enable high-fidelity spin polarization and readout through optical methods. In contrast, it is more difficult to initialize and read out the spin state of an NV$^0$ center (a spin-half system) \cite{NV0_PRL}.
Laser excitation induces charge conversion of NV centers (between NV$^-$ and NV$^0$), either by two- or one-photon ionization processes \cite{aslam2013photo,dhomkar2018charge}. 

The main finding of this work is that the internal electric field around an NV center is mainly determined by its nearby substitutional nitrogen atoms and can be modified by the excitation light, with the effect becoming more pronounced in the weak excitation regime.
It is generally believed that the extra electron of an NV$^-$ center is captured from a nearby defect, for example, another nitrogen atom (P1 center) \cite{manson2018nv}. In this scenario, the negatively charged NV$^-$ center is often accompanied by a positively charged N$^+$. 
Although the diamond is electrically neutral overall, these randomly distributed charges (NV$^-$, N$^+$, and other charge traps) lead to internal electric fields at the NV positions \cite{mittiga2018imaging}.
Under laser excitation, the charge distribution and the associated internal electric field in the diamond lattice are altered, leading to observable changes in the optical and spin properties of the remaining NV centers, as illustrated in Fig.\,\ref{fig_1}a.

We start by measuring the photoluminescence (PL) spectra of NV centers under weak laser excitation. Nanodiamonds (NDs) with ensemble NV centers are deposited on strontium titanate (or alumina) substrates and measured in a home-built confocal microscopy. The average size of NDs is about 100\,nm, with an NV concentration of about 3\,ppm (Adámas Nanotechnologies). Figure\,\ref{fig_1}b shows typical PL spectra of these NDs, measured at 8.4\,K. The ZPLs of NV$^-$ (637\, nm) and NV$^0$ (574\, nm) are well resolved. The relative amplitude of the two peaks shows pronounced dependence on the laser intensity. 
We adopt a method described by Shinei \textit{et al.} to quantitatively estimate the ratio of remaining NV$^-$ centers at different laser intensities. The method takes into account the integration of the ZPL peaks, the Huang Rhys factor, and the lifetime of the excited state of both centers (NV$^-$ and NV$^{0}$)\,\cite {NV_Ratio_2021APL}.
As summarized in Fig.\,\ref{fig_1}\,d, laser intensity plays an important role in determining the charge state of NV centers in diamond, and weak excitation leads to more NV$^-$ centers\,\cite{Charge_T1}.
Note that a similar phenomenon has been reported\,\cite{Charge_T1,ito2023optical}.

Next, we show that the laser-induced charge conversion also strongly affects the NV magnetic resonance signal. ODMR spectra of NV$^-$ centers in diamond are obtained by monitoring the photoluminescence intensity while simultaneously scanning the frequency of the applied microwave pulses\,\cite{gruber1997scanning}.
Figure\,\ref{fig_1}b shows typical ODMR spectra of the same nanodiamond, exhibiting a two-dip feature at 2.87\,GHz, the value of zero-field splitting of NV centers at cryogenic temperatures. The resonant frequencies and widths are extracted by numerical fitting, and their power dependence is summarized in Fig.\,\ref{fig_1}d. In the weak excitation region, both the splitting and the width of the ODMR spectra decrease with laser power. We measured several different NDs, all of which show a similar power dependence (see Supplemental Material for extended data)
(see also references\,\cite{van1990electric,whitehead1939measurement,chen2017near,yuan2020charge,siyushev2013optically,savinov2022diamond,roberts2019spin,giri2019selective,Lindblad1976,tan1999computational} therein) .

\begin{figure}
\includegraphics[width=0.48\textwidth]{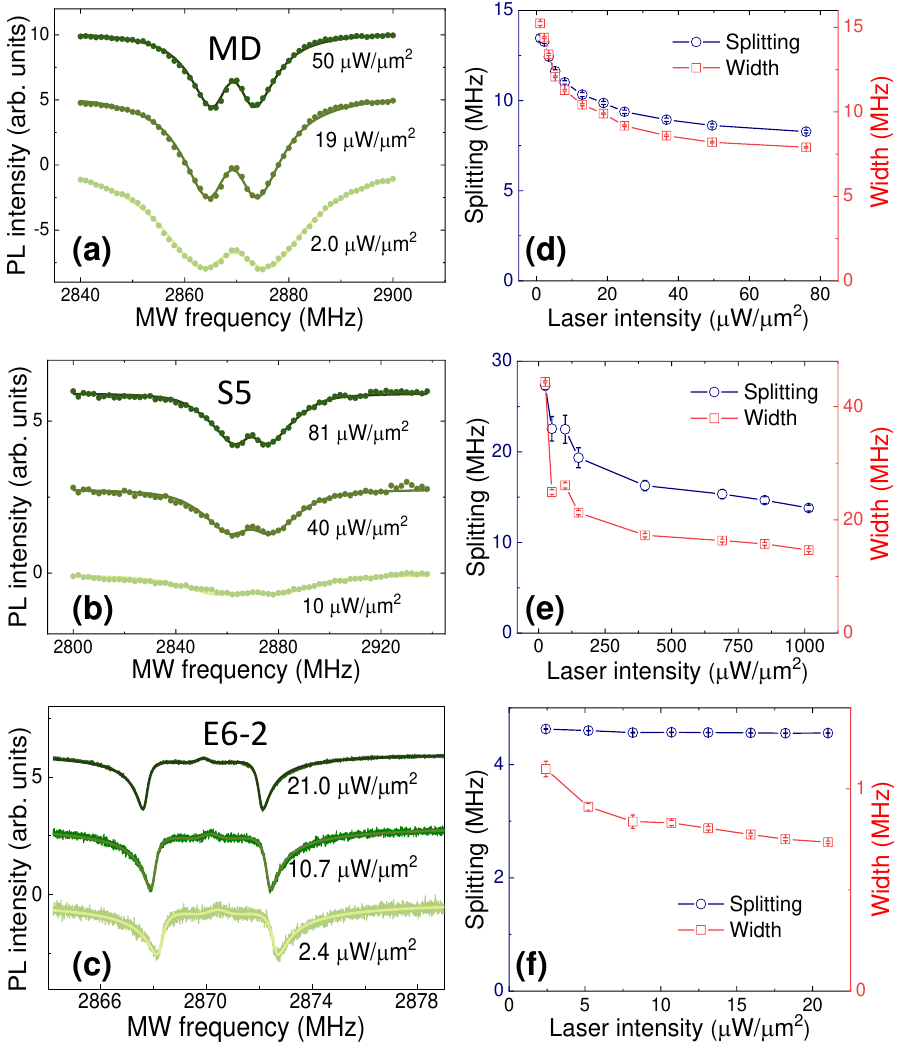}%
\caption{(a-c) ODMR spectra of NV centers in microdiamonds (MD), and bulk diamonds (S5 and E6-2) under weak laser excitation, see Table I for detailed sample information. The spectra of MD and sample S5 are measured under Earth’s magnetic field. The spectra of sample E6-2 are measured under zero magnetic field.
(d-f) Laser power dependence of the ODMR splitting and width for the (d) MD sample, (e) S5 sample, and (f) E6-2 sample. 
}
\label{fig:ODMRdata}
\end{figure}

To check whether this observation is a general effect, three other batches of diamond samples are measured.
First, ODMR spectra of ensemble NV centers in microdiamonds (MDs) and a bulk diamond (S5), with similar nitrogen concentration (about 200 ppm), are shown in Fig.\,\ref{fig:ODMRdata}\,a-b. The laser power dependence of ODMR splitting and width of these spectra, as summarized in Fig.\,\ref{fig:ODMRdata}\,d-e, are similar to that of NDs, indicating that particle size is not a key factor for this effect. 

Secondly, ODMR spectra of another bulk diamond (E6-2) with similar NV concentration but more dilute N concentration ($\sim$14\, ppm) are measured, as shown in Fig.\,\ref{fig:ODMRdata}\,c. In this measurement, pairs of coils were used to compensate for the background magnetic field in the laboratory. The main feature of these ODMR spectra is the hyperfine splitting caused by the host $^{14}$N nuclear spins. Numerical fitting considering the internal electric field and hyperfine coupling is used to extract the ODMR splitting and width. Both parameters decrease slightly with increasing laser power, as can be seen in the inset of Fig.\,\ref{fig:ODMRdata}\,f.

Summarizing, key features of NV ODMR spectra have unambiguous laser power dependence.
The effect is more pronounced in the weak excitation region and is closely related to the nitrogen concentration of the diamond sample.

\begin{table}[t]
\caption{\label{tab:sample_info}
Details of diamond samples used in this work.
}
\begin{ruledtabular}
\begin{tabular}{cccc}
\textrm{Sample}&
\textrm{size}&
\textrm{[N]}&
\textrm{[NV]}\\
\colrule
 ND            & $\sim$100 nm                                       & $\sim$ 200~$\mathrm{ppm}$      & $\sim$3\,ppm   \\     
 MD            & $\sim$ 1 $\mu$m                                     & $\sim$ 200~$\mathrm{ppm}$      & $\sim$3\,ppm     \\   
 S5            &3~$ \times $3~$ \times $0.5~mm$ ^{3} $       &$ \sim200\,\mathrm{ppm}$        & 5-40\,ppm  \\      
 E6-2            &3~$ \times $3~$ \times $0.7\,mm$ ^{3} $  &$14\,\mathrm{ppm}$           &$ \sim $3\,ppm   \\
\end{tabular}
\end{ruledtabular}
\end{table}

To understand these results quantitatively, we adapt the NV$^-$- N$^+$ pair model \cite{manson2018nv} developed by Manson \emph{et al} and pay special attention to the internal electric field around NV centers \cite{mittiga2018imaging}.
Key points of this model are:
(1) NV and N defects are randomly distributed in the diamond lattice. 
(2) Each NV center captures an electron from one of its nearby N atoms and forms an NV$^-$- N$^+$ pair. 
(3) The resonance frequencies of each NV center are modified by its local internal electric field. 
(4) Upon laser excitation, some of the NV$^-$ - N$^+$ pairs are charge neutralized, resulting in more dilute charge density and smaller internal electric fields for the remaining NV$^-$ centers.

For simplicity, we assume that an NV center always pairs with its nearby N atoms. The local electric field at the NV position contains two parts: the electric field from its paired N$^+$ ion ($E_{\rm N}$) and the dipole fields ($E_{\rm dipole}$) from all other NV$^-$-N$^+$ pairs. For the nano- and micro-diamond samples, the N concentration ($\sim$200\, ppm) is much larger than the NV concentration (3\, ppm), so the distance between different pairs is much larger than the distance between the two point charges (NV$^-$ and its nearby N$^+$). As a result, the $E_{\rm N}$ component dominates the internal electric field and the  $E_{\rm dipole}$ term can be ignored. Under these conditions, the splitting and the width of ODMR spectra of ensemble NV centers can be related to the N concentration (see Supplemental Material for detailed derivation):
\begin{equation}\label{equ:coefficient}
\begin{gathered}
S_{\rm ODMR}\approx 0.42~\rho_N^{\frac{2}{3}}~~\mathrm{MHz} / \mathrm{ppm}^{\frac{2}{3}},\\
W_{\rm ODMR}\approx 0.31~\rho_N^{\frac{2}{3}}~~\mathrm{MHz} / \mathrm{ppm}^{\frac{2}{3}}.
\end{gathered}
\end{equation}
where $\rho_N$ is the concentration of the N atoms (in ppm). It is clear that a dense N distribution leads to a large splitting and broadening feature to the ODMR spectra. When the NV and N concentrations are of the same order of magnitude, for example, in the E6-2 sample, both the $E_{\rm N}$ and $E_{\rm dipole}$ terms should be taken into account, but the nitrogen concentration dependence of ODMR parameters is similar to the former case. 

With the model in hand, we now use numerical simulations to investigate the effects of nitrogen density and laser-induced charge neutralization on the key features of ODMR spectra. NV centers and N atoms are randomly placed in a sphere of 100-nm diameter, and the total number of NV centers (and N atoms) is chosen to match the known defect concentrations.
The positions of the $i$-th NV center and the $j$-th N atom are denoted as $\left\{\vec{r}_{i}\right\}$ and $\left\{\vec{r}_{j}\right\}$, respectively.
The electric field on the $i$-th NV$^-$ center is: 
\begin{eqnarray}\label{electricfieldNV}
\vec{E^{i}}=\frac{-e}{4 \pi \epsilon_{0}\epsilon_{r}}\left [ \sum_{k\neq i}  \frac{\hat{r}_{ki}}{r_{ki}^{2}}
       -\sum_{j} \frac{\hat{r}_{ji}}{r_{ji}^{2}}\right ]\,,
\end{eqnarray}
where $\epsilon_0$ is the vacuum permittivity, $\epsilon_{r} $ = 5.7 is the relative permittivity of diamond~\cite{whitehead1939measurement},
and $ r_{ki}$ and $r_{ji}$ are the distances from the $k$-th NV$ ^{-} $ and $j$-th N$ ^{+}$ to the NV position,
$\hat{r}_{ki} $ and $ \hat{r}_{ji} $ are the corresponding unit vectors.

This electric field can be decomposed into parallel ($E^{i}_{z}$) and perpendicular ($E^{i}_{\{x, y\}}$) components relative to the NV axis, and it couples to the NV electron spin through $\Pi^{i}_{\{x, y\}}=d_{\perp} E^{i}_{\{x, y\}}$ 
and $\Pi^{i}_{z}=d_{\|} E^{i}_{z}$, with susceptibilities $\left\{d_{\|}, d_{\perp}\right\}$=
$\{0.35,17\}\,\mathrm{Hz}\cdot\mathrm{cm} / \mathrm{V}$~\cite{van1990electric,mittiga2018imaging}.
Specifically, the axial component of the electric field shifts the resonant frequency of the NV center, and the transverse component splits the resonant peaks. It is worth noting that the coefficient of the transverse component ($d_{\perp}$) is much larger than that of the axial component ($d_{\|}$), resulting in the two-dip feature of the ensemble NV ODMR spectra \cite{mittiga2018imaging}.

The NV spin Hamiltonian under the electric field is \cite{SingleNVElectricFieldSensing}:
\begin{equation}\label{nvcenterele}
\begin{gathered}
H_{i}=\left(D_{gs}+\Pi_{z}^{i}\right) S_{z}^{2}+\Pi_{x}^{i}\left(S_{y}^{2}-S_{x}^{2}\right)\\
+\Pi_{y}^{i}\left(S_{x} S_{y}+S_{y} S_{x}\right),
\end{gathered}
\end{equation}
where the $D_{\mathrm{gs}}=(2 \pi) \times 2.87~\mathrm{GHz}$ is the zero-filed splitting at room temperature,  
and $ S $ is the NV spin operator. We have ignored the hyperfine coupling to the host $^{14}$N and nearby $^{13}$C nuclear spins in the calculation, as they
are constant and do not change under laser excitation.
After the diagonalization of the Hamiltonian, the eigenvalues of each NV center and the corresponding resonant frequencies are obtained.
For a given defect concentration, the calculation has been repeated $10^5$ times, and the resonant frequencies of all NV centers are taken into account to obtain their statistical features.

\begin{figure}
\includegraphics[width=0.48\textwidth]{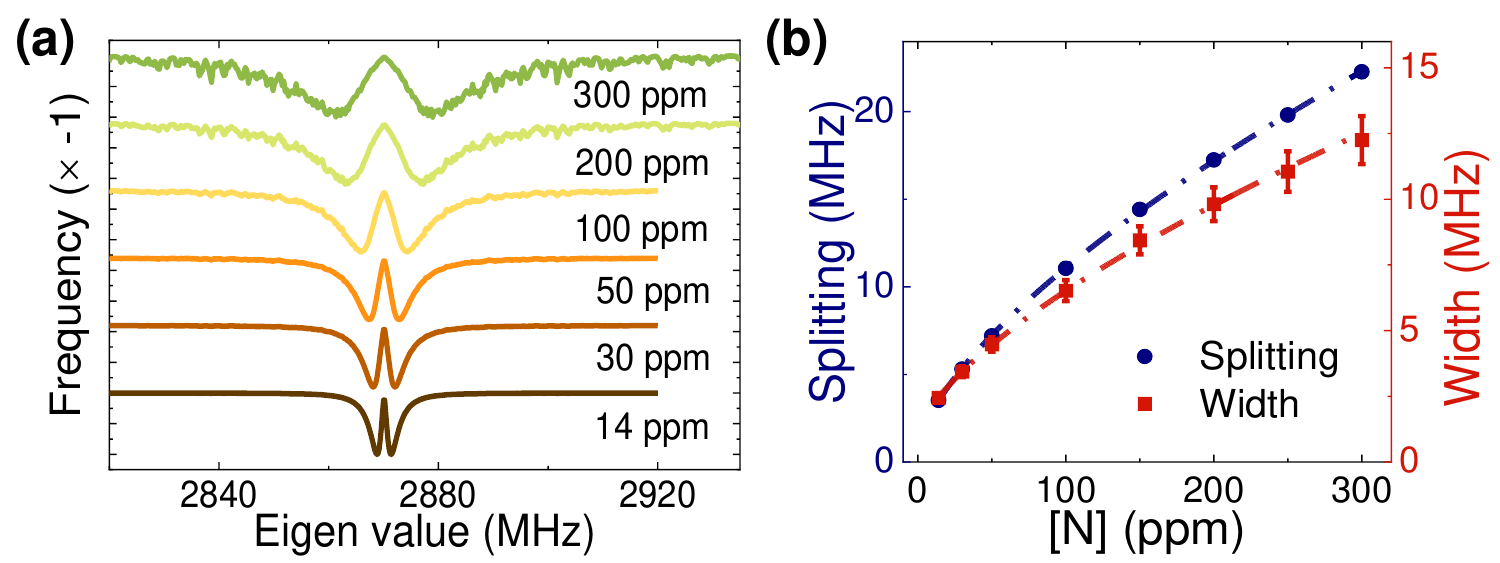}%
\caption{ Numerical results on nitrogen concentration. (a) The probability distribution of NV resonant frequency under different nitrogen concentrations ([NV$^{-}$] = 3\,ppm). (b) Splitting and width of the numerical spectra as a function of the nitrogen concentration. The dashed lines are fitting results with Eq.\,\ref{equ:coefficient}. 
Note that spin relaxation, spin dephasing, and other technical broadening factors (e.g. microwave power broadening) are not considered in this simulation.}
\label{fig_3}
\end{figure}

Figure\,\ref{fig_3}\,a presents the probability distribution of the NV resonant frequency under different N concentrations. Although intrinsic spin dephasing and technical broadening factors (e.g., microwave power broadening) have not been taken into account, these numerical spectra capture the key features of the experimentally measured ODMR spectra. The numerical results confirm that a dense N distribution leads to a large splitting and broadening to the ODMR spectra, as summarized in Fig.\,\ref{fig_3}\,b. The numerical results also fit well with the theoretical formula of Eq.\,\ref{equ:coefficient}. As an application of this result, one can use the ODMR splitting and width under zero-field to estimate the nitrogen concentration of unknown diamond samples if the light illumination is sufficiently well characterized.
From a diamond material growth and NV$^-$ center generation point of view, as a large nitrogen concentration is a prerequisite for dense NV centers, it is preferred to maximize the efficiency of nitrogen to NV center conversion and leave few nitrogen impurities (P1 centers) in the diamond lattice.

\begin{figure}[t]
\includegraphics[width=0.48\textwidth]{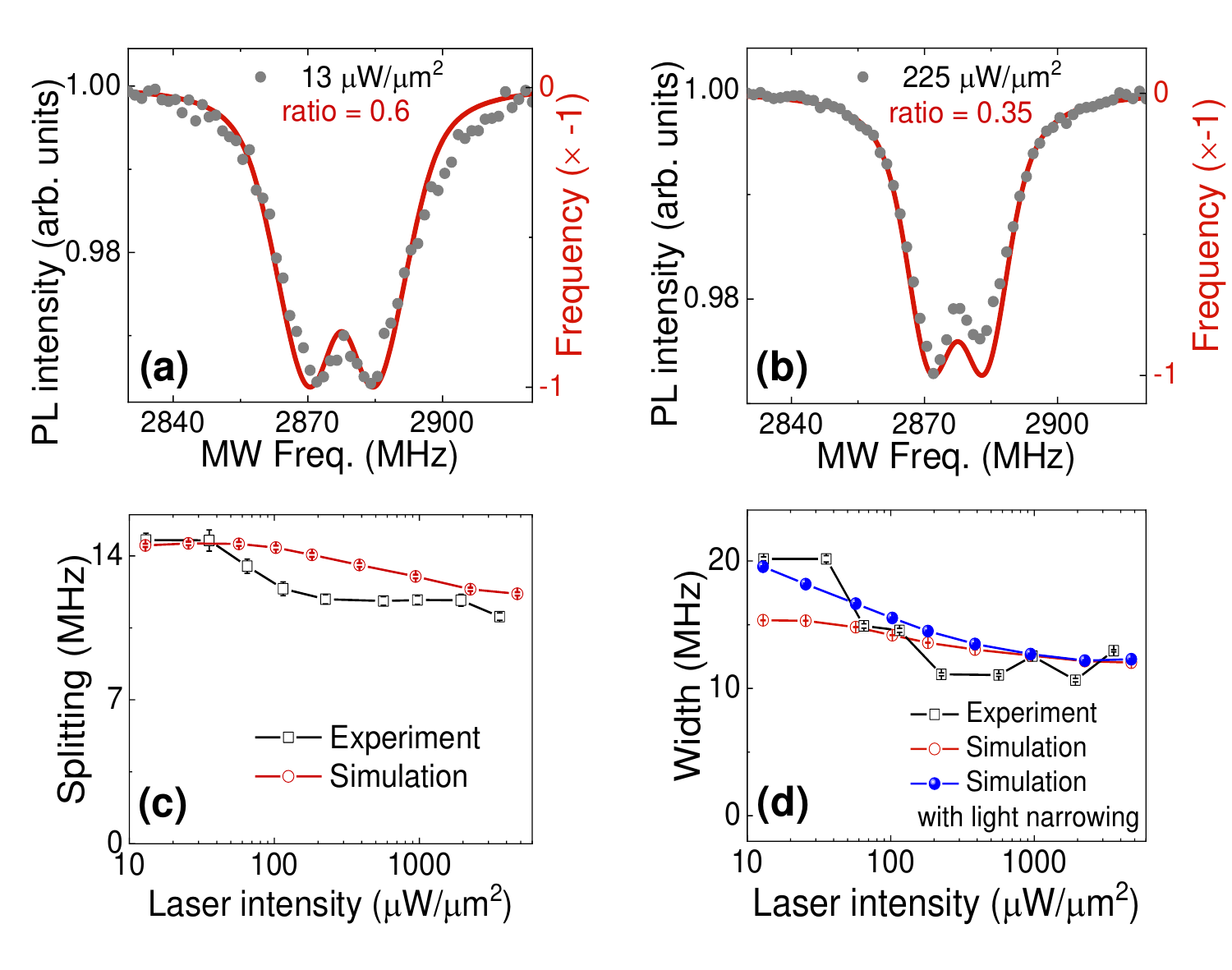}%
\caption{Comparison between numerical and experimental results. (a-b) Experimentally measured (gray dots) and numerically simulated (red lines) ODMR spectra of an ND at various laser powers. For the numerical simulation, each laser power is converted into the corresponding NV$^{-}$ ratio for the simulation. The parameters for the numerical simulation are [N] = 200\,ppm, [NV$^{-}$] = 3\,ppm, $1/T_{2}^{\star}$ = 2\,MHz and $\Omega_{R}$ = 2$\pi\times$4\,MHz. (c) Splitting of the ODMR spectra. (d) Width of the ODMR spectra. The red line are the fitting result from the simulated ODMR spectra (without light narrowing), while the blue line also considers the light narrowing effect \cite{Budker_narrowing}.}
\label{fig_4}
\end{figure}

In order to directly compare the numerical results with the experimental ODMR spectra, we incorporate the effects of NV spin dephasing and microwave power broadening in the numerical simulation. 
Considering the effect of microwave driving, the ODMR spectra are broadened by a factor proportional to ${\Omega_{R}^2}/{(\Omega_{R}^2+(2\pi\Delta)^2)}$, where $\Omega_{R}$ is the Rabi frequency and $\Delta$ is the detuning of the driving microwave.
With the following parameters: NV$^{-}$ concentration [NV$^{-}$] = 3\,ppm, nitrogen concentration [N] = 200\,ppm, $1/T_{2}^{*}$ = 2\,MHz, and $\Omega_{R}$= 2$\pi\times$4\,MHz,  the simulation results are in good agreement with the experimental ODMR spectra, as shown in Fig.\,\ref{fig_4}\,a-b. Details can be found in the Supplemental Material.

We then consider the role of laser excitation in determining the key feature of ODMR spectra in the weak excitation region. 
First-principle calculations suggest that the energy required for the NV$^0$ + N$^0$ $\Rightarrow$ NV$^-$- N$^+$ reaction is always negative\,\cite{ferrari2021nv}.
Therefore, NV$^-$- N$^+$ pairs are stable in the dark. Under laser excitation, however, the reverse process can take place, and part of the NV$^-$- N$^+$ pairs are charge neutralized. In our measured samples, this most likely takes place through a one-photon process \cite{NV_decharge}, as weak laser excitation is sufficient to induce this effect.  
The equilibrium charge state of NV centers under laser excitation can be obtained from the experimental data shown in Fig.\,\ref{fig_1}\,d. For the remaining NV$^-$ centers, we calculate their ODMR spectra using the method mentioned above.
The laser power dependence of ODMR spitting and width are shown in Fig.\,\ref{fig_4}\,c-d.

By directly comparing the numerical (red) and experimental (black) results on laser power dependence, we find that the splitting values agree well from both sides. However, there is an obvious gap in the ODMR width values. This discrepancy may be mainly caused by the effect of light narrowing. As shown in reference \cite{Budker_narrowing}, the width of ODMR spectra broadened by the microwave will be narrowed because the laser excitation effectively decreases the NV spin relaxation time ($T_1$).
To account for the effect of light narrowing, we rerun the above numerical simulation but without the effect of microwave power broadening.
The obtained ODMR widths are treated as intrinsic widths (only the effects of NV spin dephasing and charge neutralization are considered). Next, we use Equation 1 from reference \cite{Budker_narrowing} to refine the dependence of the ODMR width on the laser power. This agrees well with the experimental results, as plotted by the blue symbols in Fig.\,4\,d.

(\emph{Discussion}) 
The laser-induced charge neutralization of NV$^-$- N$^+$ pairs would affect the performance of NV-based quantum sensing.
On the one hand, the fluorescence of NV$^0$  increases the background level and decreases the spin-dependent fluorescent contrast of NV$^-$ centers.
On the other hand, the influence on the splitting and linewidth can degrade the sensitivity based on ODMR detection, especially at low magnetic fields.
Take the ODMR spectra shown in Fig.\,4\,a-b as examples: when the laser intensity increases from 13\,$\mu$W/$\mu$m$^2$ to 225\,$\mu$W/$\mu$m$^2$, the ODMR contrast decreases from 0.036 to 0.026, the ODMR width narrows from 20.2\,MHz to 11.1\,MHz and the NV fluorescence counts increase from 170\,kps (kilo-counts per second) to 2306\,kps, resulting in an improvement of the magnetic sensitivity from 377 to 78\,$\mu$T/$\sqrt{\rm Hz}$.
It is clear that higher laser intensity is preferred to increase the sensitivity of NV-based quantum sensors in the weak excitation regime.
For some applications where the laser intensity is limited by other factors, it is crucial to maintain a stable light power.
For example, in experiments with living cells, the light exposure should be less than 20\,J/cm$^2$ to avoid phototoxicity \cite{phototoxicity2017NatMeth}.
Assuming an experimental integration time of 1 minute and disregarding other factors (e.g. wavelength), the safe laser intensity is approximately 3.0\,nW/$\mu$m$^2$. Therefore, it is necessary to maintain a stable laser power in bio-sensing applications of diamond NV centers.

The contribution of internal electric fields becomes more significant at zero and low magnetic fields, so applications of NV-based quantum sensors under these circumstances should pay particular attention to this effect.  
For example, zero-field magnetometry\,\cite{zheng2019zero, 2022PRApp_Zero_Field_Sesning, 2022PRR_Zero_Field_Sesning}, zero- and ultralow-field nuclear magnetic resonance\,\cite{ZULF-NMR}, magnetoneurography and magnetomyography\,\cite{NV_Magnetoneurography}, temperature sensing and intracellular sensing\,\cite{2013T_sensing,2020T_sensing, 2020T_sensing,choi2020PNAS}, and so on. Note that, although the light-induced charge state conversion is a general effect, an external magnetic field could be applied to suppress the electric-field coupling and mitigate the influence on magnetic sensitivity. 
We also measured the laser power dependence of ODMR width under an external magnetic field about 58\,G. 
As shown in Fig.\,S6\,c, only a slight narrowing effect has been observed, indicating a significant suppression on the coupling of the internal electric field by a large Zeeman splitting.
It is worth noting that the dependence of ODMR splitting on laser intensity at (near) zero magnetic field cannot be explained by the light narrowing effect alone.

In our numerical simulation, we made the simplified assumption that an NV center pairs only with its nearest nitrogen atom. In fact, with multiple N atoms available, an NV center could also pair with other nitrogen atoms. In addition, the pair "partner" of a particular NV center may change over time.  
Nevertheless, the good agreement between the simulation and experimental results indicates that the simplified model captures the key features of laser-induced charge neutralization of NV$^-$- N$^+$ pairs and the changes of local electric field environment.

In summary, we experimentally observe unambiguous laser-power dependence of (near) zero-field ODMR spectra of NV ensembles in nano-, micro-, and bulk diamonds. In the weak-excitation regime, both the width and the splitting of the ODMR lines decrease with increasing laser power, which is accompanied by a reversible change of the height of the NV$^-$ ZPL peak. We use the NV$^-$- N$^+$ pair model and perform numerical simulations to verify the effect of laser-induced charge neutralization. It turns out that both the laser power and the nitrogen-concentration dependence can be well understood with this simple model.
These results will be useful in diamond-based quantum sensing applications, for example, probing biological intracellular signals and in studies of temperature-sensitive thin-film materials. 
\section*{Acknowledgements}
The authors acknowledge helpful discussions with Neil Manson and Till Lenz. 
This work was supported by the Natural Science Foundation of Beijing, China (Grant No. Z200009), the National Key Research and Development Program of China (Grants No. 2019YFA0308100), the Chinese Academy of Sciences (Grant Nos. YJKYYQ20190082, XDB28030000, XDB33000000), and the National Natural Science Foundation of China (Grant Nos.\,11974020, 12022509, 11934018, T2121001). 
Q. L. and R.-B.L. acknowledge the funding support from Hong Kong Research Grants Council - Collaborative Research Fund under project no. C4007-19G. The work of DB was supported by the European Commission’s Horizon Europe Framework Program under the Research and Innovation Action MUQUABIS GA No.\,101070546. The work of HZ was supported by Beijing Natural Science Foundation No.\,L233021.
\bibliographystyle{apsrev4-2-2}
\bibliography{literature.bib}

\end{document}


\title{Supplemental Material: Optically Detected Magnetic Resonance with Nitrogen-Vacancy Centers in Diamond under Weak Laser Excitation}

\author{Yong-Hong Yu}
\author{Rui-Zhi Zhang}
\author{Yue Xu}
\affiliation{Beijing National Laboratory for Condensed Matter Physics, Institute of Physics, Chinese Academy of Sciences, Beijing 100190, China}
\affiliation{School of Physical Sciences, University of Chinese Academy of Sciences, Beijing 100049, China}

\author{Huijie Zheng}
\email{hjzheng@iphy.ac.cn}
\affiliation{Beijing National Laboratory for Condensed Matter Physics, Institute of Physics, Chinese Academy of Sciences, Beijing 100190, China}
\affiliation{Songshan Lake Materials Laboratory, Dongguan, Guangdong 523808, China}

\author{Quan Li}
\author{Ren-Bao Liu}
\affiliation{Department of Physics, Centre for Quantum Coherence, and The Hong Kong Institute of Quantum
Information Science and Technology, The Chinese University of Hong Kong, New Territories, Hong Kong, China}

\author{Xin-Yu Pan}
\affiliation{Beijing National Laboratory for Condensed Matter Physics, Institute of Physics, Chinese Academy of Sciences, Beijing 100190, China}
\affiliation{Songshan Lake Materials Laboratory, Dongguan, Guangdong 523808, China}
\affiliation{CAS Center of Excellence in Topological Quantum Computation, Beijing 100190, China}

\author{Dmitry Budker}
\affiliation{Johannes Gutenberg-Universit{\"a}t Mainz, 55128 Mainz, Germany}
\affiliation{Helmholtz-Institut, GSI Helmholtzzentrum f{\"u}r Schwerionenforschung, 55128 Mainz, Germany}
\affiliation{Department of Physics, University of California, Berkeley, California 94720, USA}

\author{Gang-Qin Liu}
\email{gqliu@iphy.ac.cn}
\affiliation{Beijing National Laboratory for Condensed Matter Physics, Institute of Physics, Chinese Academy of Sciences, Beijing 100190, China}
\affiliation{Songshan Lake Materials Laboratory, Dongguan, Guangdong 523808, China}
\affiliation{CAS Center of Excellence in Topological Quantum Computation, Beijing 100190, China}
\maketitle
\tableofcontents
\pagebreak 
\section{Experimental setup and sample preparation}
The ODMR and PL measurements of nanodiamonds, microdiamonds, and sample S5 (bulk diamond) are carried out on home-built confocal microscopies. The light source is a continuous-wave 532\, nm laser (CNI, MLL-III-532-150 mW), and the laser power can be adjusted by neutral density filters (Thorlabs) directly after the laser. An objective with a numerical aperture of 0.9 is used to focus the laser beam on the diamond samples and also to collect the NV fluorescence. The focal point has an estimated waist of 1\,$\mu$m.
Confocal images of the diamond samples are obtained by scanning either the sample holder (with piezoelectric stages) or the incident laser beam (with galvo mirrors).

ODMR spectra are measured by recording the fluorescence (between 650\, nm with 800\, nm) photon counts of the NV centers as a function of the driving microwave (MW) frequency. The microwaves are generated by signal generators (Rohde \& Schwarz SMIQ06B, or Keysight N5183B), controlled in shape by an RF switch (Mini-Circuits, ZASWA-2-50DR+), then amplified by a high-power amplifier (Mini-Circuits, ZHL-16W-43-s+), and delivered to the antenna (a twenty micrometer thick silver wire) on the sample holder. No external magnetic field is applied in all the ODMR measurements. For the E6-2 sample, a set of three-dimensional Helmholtz coils is used to compensate for the geomagnetic field, as the splitting due to the internal electric field is small (less than 1\, MHz).

\begin{figure}[H]
\centering
\includegraphics[width=0.8\textwidth]{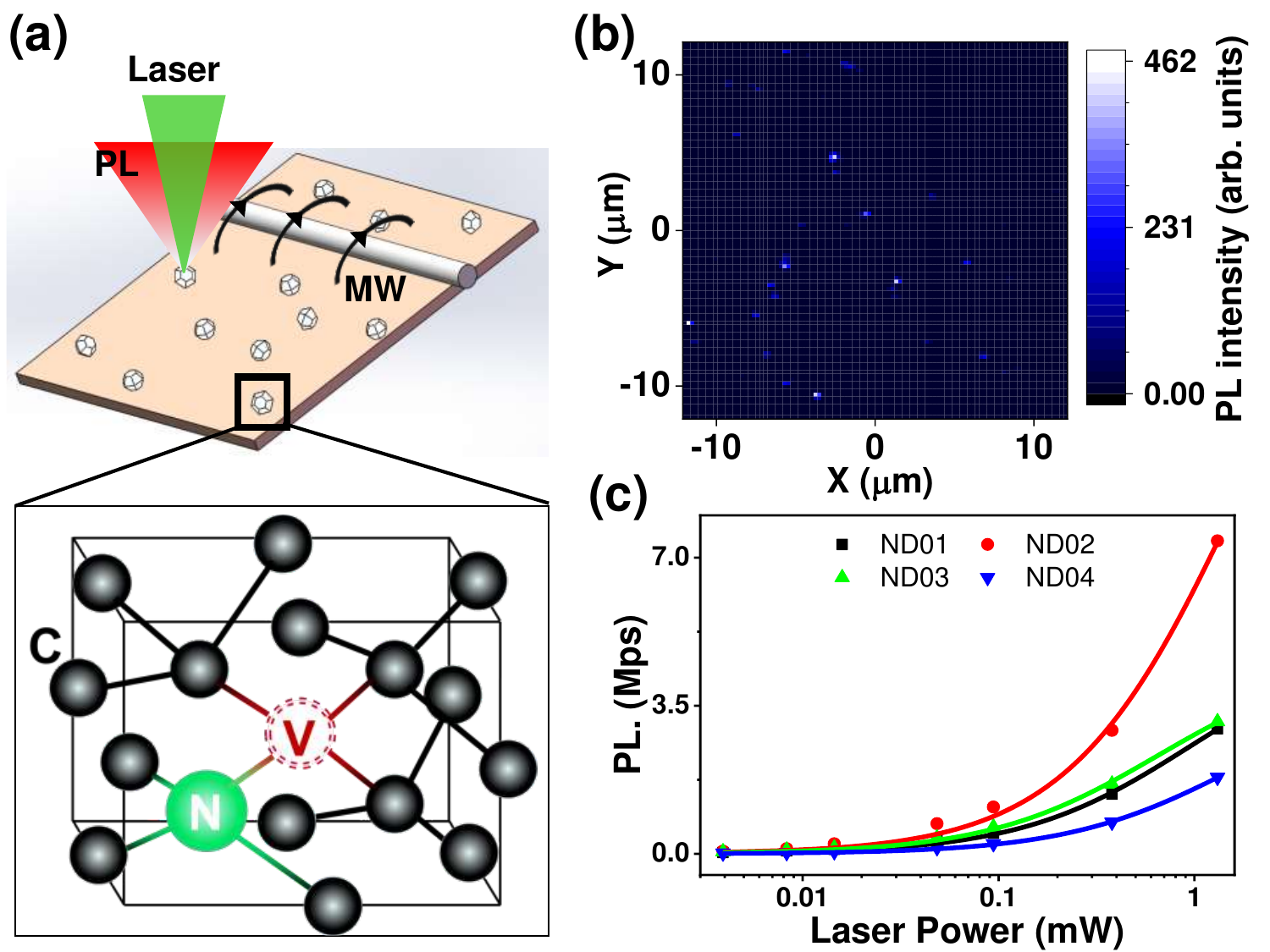}
\caption{(a) Schematic representation of the experiment.
(b) Confocal image of nanodiamonds (NDs) on alumina substrate.
(c) Laser power dependence of NV fluorescence counts, symbols are experimental data, solid lines are fit to a function of the form P$ _{\rm{PL.}} $ = kP /(P + P$ _{s} $)~\cite{Budker_narrowing}. }
\label{Fig_S1}
\end{figure}

 Nanodiamonds (NDs) and microdiamonds (MDs) containing ensemble NV centers are produced by Adámas Nanotechnologies. These diamond particles are supplied in deionized water at a mass concentration of 1 mg/mL. After dilution with isopropanol and ultrasonic treatment, a drop of 5\,$\mu$L ND isopropanol solution (2.5 $\mu$g/mL) is transferred to a substrate (STO or Al$_{2}$O$_{3}$) using a pipette. After volatilization of the isopropanol, nanodiamonds are deposited on the substrate, as illustrated in Fig.~\ref{Fig_S1}~(a).
Fig.~\ref{Fig_S1}~(b) shows a typical confocal scanning image of the nanodiamonds sample. 
To estimate the laser excitation efficiency of the confocal system, we measured the laser power dependence of the NV fluorescence counts, and the results are shown in Fig.~\ref{Fig_S1}~(c). 
The saturation power is more than 1 mW of  (5000 $\mu$W/$\mu$m$^2$ in intensity). 

\section{Theoretical model and numerical fitting}\label{Theory}
We adopt the NV$^-$- N$^+$ pair model \cite{manson2018nv,mittiga2018imaging} to explain the observed laser power dependence of  ODMR spectra. The key point of our model is that each NV center captures an electron from one of its nearby nitrogen atoms (N) and forms an NV$^-$- N$^+$ pair. The randomly distributed charges in the diamond lattice (NV$^-$-, N$^+$, and other charge traps) bring a local electric field to each NV$^-$ center, leading to observable effects on the PL and ODMR signals. For a given local electric field at the position of the NV$^{-}$ center, $\vec{E}$, couples with the spin of the NV$^{-}$ center via the Hamiltonian:
\begin{equation}\label{probability14}
\hat{H}=\left(D_{g s}+\Pi_z\right) S_z^2+\Pi_x\left(S_y^2-S_x^2\right)
    +\Pi_y\left(S_x S_y+S_y S_x\right),
\end{equation}
where $D_{gs} = 2870~\mathrm{MHz}$ is the zero-field splitting of an NV$^{-}$ center, $S$ are the electronic spin-1 operators of the NV$^{-}$ center, 
The terms $\Pi_{\{x, y\}}=d_{\perp} E_{\{x, y\}}$ 
and $\Pi_{z}=d_{\|} E_{z}$ characterize the coupling between the NV$ ^{-} $'s electronic spin and its surrounding electric field, $\vec{E}$,  
with susceptibilities $\left\{d_{\|}, d_{\perp}\right\}$=
$\{0.35,17\}~\mathrm{Hz}\cdot\mathrm{cm} / \mathrm{V}$\,\cite{van1990electric}. We ignore the parallel component of the electric field in the following discussion, as its contribution is about two orders of magnitude smaller than the perpendicular components.
The eigenvalues of this Hamiltonian are $\varepsilon^{\pm}=D_{gs} \pm \Pi_{\perp}$ and $\varepsilon^{0}=0$.

In this section, we first derive the probability distribution of the distance between NV$^-$- N$^+$ pairs,  and then use this result to calculate the internal electric field of NV$^-$ centers in diamond. We obtain the formula to fit the experimental ODMR spectra at the end of this section.
\subsection{Probability distribution of the nearest NV$^-$ and N$^{+}$ pairs}
In our model, the probability distribution of the length of an NV$^-$ and N$^+$ pair is actually the distance between the NV and its nearest nitrogen neighbor, which is based on the assumption that an NV center always pairs preferentially with the nearest one. This subsection provides a mathematical model for the probability of finding nitrogen atoms at a certain distance ($r$) from an NV center. To simplify the derivation, we suppose an NV center is located at the center of a diamond sphere. There are total $N$ nitrogen atoms in the sphere and its volume is $V$. $P(r)dr$ is the probability of finding a nitrogen atom at a distance between $r$ and $r+dr$ from the NV center.
As the nitrogen atoms are randomly distributed in the diamond lattice, 
$P(r)dr$ is proportional to the volume of a thin spherical shell with thickness $dr$ and radius $r$, which is $4\pi r^2dr/V$. 
The probability that a nitrogen atom is located at a distance greater than $r$ is $\left(V-\frac{4}{3}\pi r^{3}\right)/V$.
Using these formulae, we can write $ P(r)dr $ as:
\begin{equation}
P(r) d r=\sum_{i=1}^N C_N^i\left(\frac{4 \pi r^2 d r}{V}\right)^i\left(\frac{V-\frac{4}{3} \pi r^3}{V}\right)^{N-i},
\end{equation}
where $ C_{N}^{i} $ is the binomial coefficient, representing the number of ways to choose $i$ nitrogen atoms from a total number $ N $, which is calculated as follows,
\begin{equation}
N= n^{'}_0 \cdot\rho_N \cdot V ,
\end{equation}
where $\rho_{N}$ is the nitrogen concentration in ppm, $n^{'}_0 = 1.76\times 10^{-4}$ (ppm$\cdot$nm$^3$)$^{-1}$ is the factor relating the number density (in ppm) to the volume density. Now we take $n_0 =1/2\cdot n^{'}_0$ for considering the number of atoms in a primitive cell. 
For a normal ensemble NV diamond sample, $ N\gg 1$ and $ \rho_N\ll 1$. Therefore, we can  
simplify the expression of $ P (r) $ as:
\begin{equation}\label{probability02}
P(r)=8 \pi \rho_N n_0 r^2 \exp \left(-\frac{8}{3} \pi \rho_N n_0 r^3\right).
\end{equation}
The average distance between the NV center and its nearest nitrogen atom can be calculated as :
\begin{equation}
\langle r\rangle=\int_0^{\infty} r P(r) d r=\left(\frac{3}{8\pi}\right)^{1/3}\Gamma\left [4/3\right ] \left(\rho_N n_0\right)^{-1/3},
\end{equation}
where $\Gamma[z+1]$ is the Gamma function defined as $\Gamma[z+1] = \int_0^{\infty} \mu^{z}\exp{(-\mu)} ~d\mu$.

In addition to the average distance, we can calculate the standard deviation of the distribution. 
The standard deviation, $\sigma_{r}$, is given by:
\begin{equation}
\sigma_{r}=\sqrt{\langle r^2\rangle-\langle r\rangle^2}
=\left(\frac{3}{8\pi}\right)^{1/3}
\sqrt{\Gamma[5/3]-\Gamma[4/3]^{2}}\,\,(\rho_N n_0)^{-1/3}.
\end{equation}

\subsection{Analytical derivation of the ODMR spectra}
For the case of $\rho_N \gg \rho_{N V}$, that is, when the nitrogen concentration is much larger than that of the NV concentration, the distance between the paired $\mathrm{NV}^{-}$ and $\mathrm{N}^{+}$ is much smaller than the distance between two neighboring pairs.
Therefore, the local electric field at the $\mathrm{NV}^{-}$ position is dominated by the field generated by the paired $\mathrm{N}^{+}$ atom ($E_{\rm{N}}$), and the probability distribution of $E_N$ can be derived using the differential relation $F_1\left(E_N\right) d E_N=P_1\left(r\right) d r$:
\begin{equation}\label{probability03}
F_1\left(E_{\rm{N}}\right)=-\frac{4 \pi \rho_{{\rm{N}}} n_0 \alpha^3}{E_{\rm{N}}^{5/2}}\exp \left(-\frac{8\pi \rho_{\rm{N}} n_0 \alpha^3}{3E_{\rm{N}}^{3/2}}\right),
\end{equation}
where $\alpha=\left(\frac{e}{4 \pi \varepsilon_0 \varepsilon_r}\right)^{1 / 2}$, $\epsilon_{0}$ is the vacuum permittivity, 
~$\epsilon_{r} $ = 5.7 is the relative permittivity of diamond~\cite{whitehead1939measurement}.

The relationship between the resonant frequency of an $\mathrm{NV}^{-}$ center and its local electric field is:
\begin{equation}\label{probability04}
\left|v^{\prime}-D_{gs}\right|= d_{\perp}E_{\perp}=\frac{\sqrt{6}}{3} d_{\perp}E_{\rm{N}}.
\end{equation}
The coefficient $\frac{\sqrt{6}}{3}$ is used to re-scale the strength of the electric field considering the random orientation of the internal electric field.
Using this relation, the distribution of $F_1\left(E_N\right)$ yields the distribution of eigenenergies, which corresponds to the ODMR spectrum of:
\begin{equation}\label{probability05}
g_1(v)=-\frac{4\pi \rho_{\rm{N}} n_{0} \alpha^{\prime3}}{\left|v-D_{gs}\right|^{5/2} }
\exp \left(-\frac{8\pi \rho_{{\rm{N}}} n_{0} \alpha^{\prime3}}{3\left|v-D_{gs}\right|^{3/2}}\right),
\end{equation}
where $ \alpha^{\prime}=\left(\frac{\sqrt{6}}{3}d_{\perp}\right)^{1/2}\alpha $.
It is easy to derive the splitting ($S_1$) and the width ($W_1$) of $g_1(v)$, which are:
\begin{equation}\label{probability06}
\begin{gathered}
S_1=2\left(\frac{8}{5} \pi n_0 \alpha^{\prime3}\right)^{2/3} \rho_{\rm{N}}^{2/3},\\
W_1=1.463\left(\frac{8}{5} \pi n_0 \alpha^{\prime3}\right)^{2/3} \rho_{\rm{N}}^{2/3}.
\end{gathered}
\end{equation}
Considering the relevant parameters of NV$^{-}$ center, 
we can get:
\begin{equation}\label{probability07}
\begin{gathered}
S_1=0.42~\rho_{\rm{N}}^{\frac{2}{3}}~~\mathrm{MHz} / \mathrm{ppm}^{\frac{2}{3}},\\
W_1=0.31~\rho_{\rm{N}}^{\frac{2}{3}}~~\mathrm{MHz} / \mathrm{ppm}^{\frac{2}{3}}.
\end{gathered}
\end{equation} 
The above derivation shows that when $\rho_{\rm{N}} \gg \rho_{\rm{NV}}$, 
the splitting and width of the NV ODMR spectra are mainly determined by the nitrogen concentration and have little dependence on the $\mathrm{NV}^{-}$ concentration.
Note that in the present approach, an NV$^-$ center tends to pair with its nearby N$^+$. This differs from the approximation used in\,\cite{mittiga2018imaging} where random distributions over the crystal were assumed.

\subsection{Fitting function for the ODMR spectra}
For the E6-2 sample, $\rho_{\rm{NV}}$ is close to $\rho_{\rm{N}}$, and the distance between two $\mathrm{NV}^{-}$ centers is comparable to that between $\mathrm{NV}^{-}$ and the nearest $\mathrm{N}^{+}$.
In this situation, the electric field generated by nearby $\mathrm{NV}^{-}$-$\mathrm{N}^{+}$ pairs ($E_{dipole}$) should also be taken into account.
The nearest pair can be considered as a dipole with a distance $r_2$ from the NV center, and the probability distribution of $E_{dipole}$ can be derived using the differential relation $F_2\left(E_{dipole}\right) d E_{dipole}=P_2\left(r_2\right) d r_2$:
\begin{equation}\label{probability08}
F_2\left(E_{dipole}\right)=-\frac{8\pi\rho_{\rm{NV}}n_{0}\beta^{3}}{3E_{dipole}^{2}\rho_{{\rm{N}}}^{1/3}} \exp\left(-\frac{8\pi\rho_{\rm{NV}}n_{0} \beta^{3} }{3E_{{dipole}}\rho_{\rm{N}}^{1/3}} \right),
\end{equation}
where $\beta=\left(\frac{2 \Gamma(4 / 3) e}{4 \pi \varepsilon_0 \varepsilon_r}\right)^{1 / 3}\left(\frac{8}{3} \pi n_0\right)^{-1 / 9}$.
Similarly, the distribution of eigenenergy derived from $F_2\left(E_{dipole}\right)$ is:
\begin{equation}\label{probability09}
g_2(v)=-\frac{8\pi \rho_{\rm{NV}} n_0 \beta^{\prime3}}{3\left|v-D_{gs}\right|^{2}\rho_{\rm{N}}^{1/3}} 
                  \exp\left(-\frac{8\pi \rho_{\rm{NV}} n_0 \beta^{\prime3} }{3\left|v-D_{gs}\right|\rho_{\rm{N}}^{1/3}}\right),
\end{equation}
where $ \beta^{\prime}=\left(\frac{\sqrt{6}}{3} d_{\perp}\right)^{1/3}\beta$.
The total electric field $E_{total}$ is the vector sum of $E_N$ and $E_{dipole}$, 
and can be expressed as:
\begin{equation}
E_{total}=\sqrt{E_{\rm{N}}^2+E_{dipole}^2+2 E_{\rm{N}} E_{dipole} \cos \Omega},
\end{equation}
where $\Omega$ is the angle between these two vectors.
Thus, we have
\begin{equation}\label{probability10}
F(E)=\frac{d}{d E} \int_0^{2 \pi}\left(~~\iint\limits_{0 \leq E_{total} \leq E}F_1 F_2 d E_{\rm{N}} d E_{dipole}\right) \frac{d \Omega}{2 \pi}
\sim F_1(E) * F_2(E),
\end{equation}
where $F_1(E)$ is the probability distribution of $E_{\rm{N}}$, $F_2(E)$ is the probability distribution of E$_{dipole}$. 
Accordingly, the total $g(v)$ (ignoring laser broadening and natural broadening) has a similar format:
\begin{equation}\label{probability11}
g(v) \sim g_1(v) * g_2(v).
\end{equation}

If we look at Eq.(\ref{probability05}) and Eq.(\ref{probability09}), we can see that $g_1(v)$ and $g_2(v)$ have the same mathematical form:
\begin{equation}\label{probability12}
g_i\left(\Pi_{\perp}\right) \propto-\Pi_{\perp}^{-n_{i}} \exp \left(-k_{i}\Pi_{\perp}^{-n_{i}+1}\right).
\end{equation}
Here, $2~\Pi_{\perp}=2~d_{\perp} E_{\perp}$ indicates the splitting due to the electric field.
After the convolution $ g_{1} $ and $ g_{2} $, the result spectrum should maintain the original form:
\begin{equation}\label{probability13}
g\left(\Pi_{\perp}\right)\propto-\Pi_{\perp}^{-n} \exp \left(-k \Pi_{\perp}^{-n+1}\right).
\end{equation}

For a single NV$^{-}$ center, the ODMR spectrum is determined by both its intrinsic (limited by spin dephasing time) and experimental broadening (power broadening of laser and MW):
\begin{eqnarray}\label{probability15}
s\left(v, \Pi_{\perp}\right)=
\left(\frac{C}{1+\left(\dfrac{v-\varepsilon^{+}}{\delta}\right)^2}+
      \frac{C}{1+\left(\dfrac{v-\varepsilon^{-}}{\delta}\right)^2}\right).
\end{eqnarray}
Here, $C$ is the contrast of the resonant dips, and $\delta$ is the full width at half the maximum of the spectrum.
The experimental ODMR spectra of ensemble NV centers are then denoted as:
\begin{eqnarray}\label{probability16}
G(v)&=&\int_0^{+\infty} s\left(v, \Pi_{\perp}\right) g\left(\Pi_{\perp}\right) d \Pi_{\perp}+ C_{0}\nonumber\\
&=&C(\int_0^{+\infty} \frac{\Pi_{\perp}^{-n} \exp \left(-k \Pi_{\perp}^{-n+1}\right)}{1+\left(\frac{v-D_{g s}-\Pi_{\perp}}{\delta}\right)^2} d \Pi_{\perp}+\int_0^{+\infty} \frac{\Pi_{\perp}^{-n} \exp \left(-k \Pi_{\perp}^{-n+1}\right)}{1+\left(\frac{v-D_{g s}+\Pi_{\perp}}{\delta}\right)^2} d \Pi_{\perp})+C_{0},
\end{eqnarray}
where $ C_{0} $ is a offset.

Considering the nuclear spin of the host $^{14}\mathrm{N}$, we add the hyperfine term to the original Hamiltonian:
\begin{eqnarray}\label{probability17}
\hat{H}=\left(D_{g s}+\Pi_z\right) S_z^2+\Pi_x\left(S_y^2-S_x^2\right)+\Pi_y\left(S_x S_y+S_y S_x\right) +A_{z z} I_z S_z,
\end{eqnarray}
here, $I$ is the nuclear spin operator. 
The energy level is further splitted by: $\varepsilon_1^{ \pm}=D_{g s} \pm \Pi_{\perp},  \varepsilon_2^{ \pm}=D_{g s} \pm \sqrt{\Pi_{\perp}^2+A_{z z}^2}$.
The formula of ODMR spectra becomes:
\begin{eqnarray}\label{probability18}
 G(v)&=&C_1(\int_0^{+\infty} \frac{\Pi_{\perp}^{-n} \exp \left(-k_1 \Pi_{\perp}^{-n+1}\right)}{1+\left(\frac{v-D_{g s}-\Pi_{\perp}}{\delta_1}\right)^2} d \Pi_{\perp}+\int_0^{+\infty} \frac{\Pi_{\perp}^{-n} \exp \left(-k_1 \Pi_{\perp}^{-n+1}\right)}{1+\left(\frac{v-D_{g s}+\Pi_{\perp}}{\delta_1}\right)^2} d \Pi_{\perp})\\\nonumber
 &+&C_2(\int_0^{+\infty} \frac{\Pi_{\perp}^{-m} \exp \left(-k_2 \Pi_{\perp}^{-m+1}\right)}{1+\left(\frac{v-D_{g s}-\sqrt{\Pi_{\perp}^2+A_{z z}^2}}{\delta_2}\right)^2} d \Pi_{\perp}+\int_0^{+\infty} \frac{\Pi_{\perp}^{-m} \exp \left(-k_2 \Pi_{\perp}^{-m+1}\right)}{1+\left(\frac{v-D_{g s}+\sqrt{\Pi_{\perp}^2+A_{z z}^2}}{\delta_2}\right)^2} d \Pi_{\perp})+ C_{B}.
\end{eqnarray}

\section{Numerical simulation of ensemble NV ODMR}
\label{sec:OdmrNitrogen}
In this section, taking nitrogen concentration and laser power density as key tuning parameters, we use numerical simulation to reproduce the ODMR spectra of ensemble NV centers in diamond. 
First, we focus on the nitrogen concentration and ignore other effects, which results in Figure 3 of the main text.
Next, for a given nitrogen concentration, we further incorporate the effects of intrinsic spin dephasing and microwave power broadening. With these factors, the simulated spectra agree well with the experimentally measured ODMR spectra. 
Science laser excitation triggers charge conversion of NV centers and the ratio of remaining NV$^{-}$ centers can be estimated from the low-temperature PL spectra, the effect of laser excitation can be simulated by setting an appropriate initial NV$^{-}$ density. 
Finally, by solving a master equation that includes the optical excitation and the charge conversion rates of both NV$^{-}$ and NV$^{0}$, the equilibrium charge state of NV centers under laser excitation is simulated.
\subsection{Nitrogen concentration dependence}
The simulation is carried out for a diamond sphere with a diameter of 100\, nm. The [NV$^{-}$] concentration is set to 3\, ppm. 
The individual steps are listed below:
\begin{enumerate}
\item Calculating the number of NV$^{-}$ and N in the given volume. In this simulation, there are 7.93$\times$10$^{7}$ carbon atoms, 7.93$\times$10$^{3}$ nitrogen (N), and 238 NV$^{-}$ centers.

\item Distribution of N and NV$^{-}$ centers in the lattice. The NV$^{-}$ centers and N atoms are randomly distributed over the 100-nm sphere. It is worth noting that if one defect coincides in the same primitive cell with another, the point is removed and recalculated until the total number of defects are generated in discrete positions.

\item Formation of NV$^{-}$-N$^{+}$ pairs. We assume that an NV$^{-}$ center always pairs with its nearest N atom and forms an NV$^{-}$-N$^{+}$ pair. We denote the position of the $i$-th NV$ ^{-} $ and $j$-th N$ ^{+} $ in by $\left\{\vec{r}_{i}\right\}$ and $\left\{\vec{r}_{j}\right\}$, respectively.

\item Calculation of the internal electric field. For each NV$^{-}$ center, its local electric field is the vector sum of the electric fields of all other charges:
\begin{eqnarray}\label{electricfieldNV}
\vec{E^{i}}=\sum_{k\neq i} \frac{-e}{4 \pi \epsilon_{0} \epsilon_{r}} \frac{\hat{r}_{ki}}{r_{ki}^{2}}
       +\sum_{j} \frac{e}{4 \pi \epsilon_{0} \epsilon_{r}} \frac{\hat{r}_{ji}}{r_{ji}^{2}}\,,
\end{eqnarray}
where,~$ r_{ki} $ and $ r_{ji} $ are the distances from the $k$-th NV$ ^{-} $ and $j$-th N$ ^{+} $  
to the $i$-th NV$ ^{-} $, respectively; 
$ \hat{r}_{ki} $, and $ \hat{r}_{ji} $ are the unit vectors for
~$ r_{ki} $ and $ r_{ji} $, respectively.
\item Diagonalization of the Hamiltonian. The local electric field at the position of the $i$-th NV$ ^{-} $ center is coupled to its spin via the Hamiltonian given by Eq.\,\eqref{probability14}.
The Hamiltonian of each NV$ ^{-} $ center is diagonalized, and its three eigenvalues are recorded.  The resonant frequencies of each NV$^{-} $ center are then calculated.

\item Repeat the 2-5th steps 10,000 times.

\item Distribution of the resonant frequency and normalization. By aggregating all the resonant frequencies, their distribution can be obtained. 
The distribution of the resonant frequency is normalized and its negation is plotted in the main text, as the experimentally measured ODMR spectra have negative contrast.
\end{enumerate}
\begin{figure}
\centering
\includegraphics[width=0.5\textwidth]{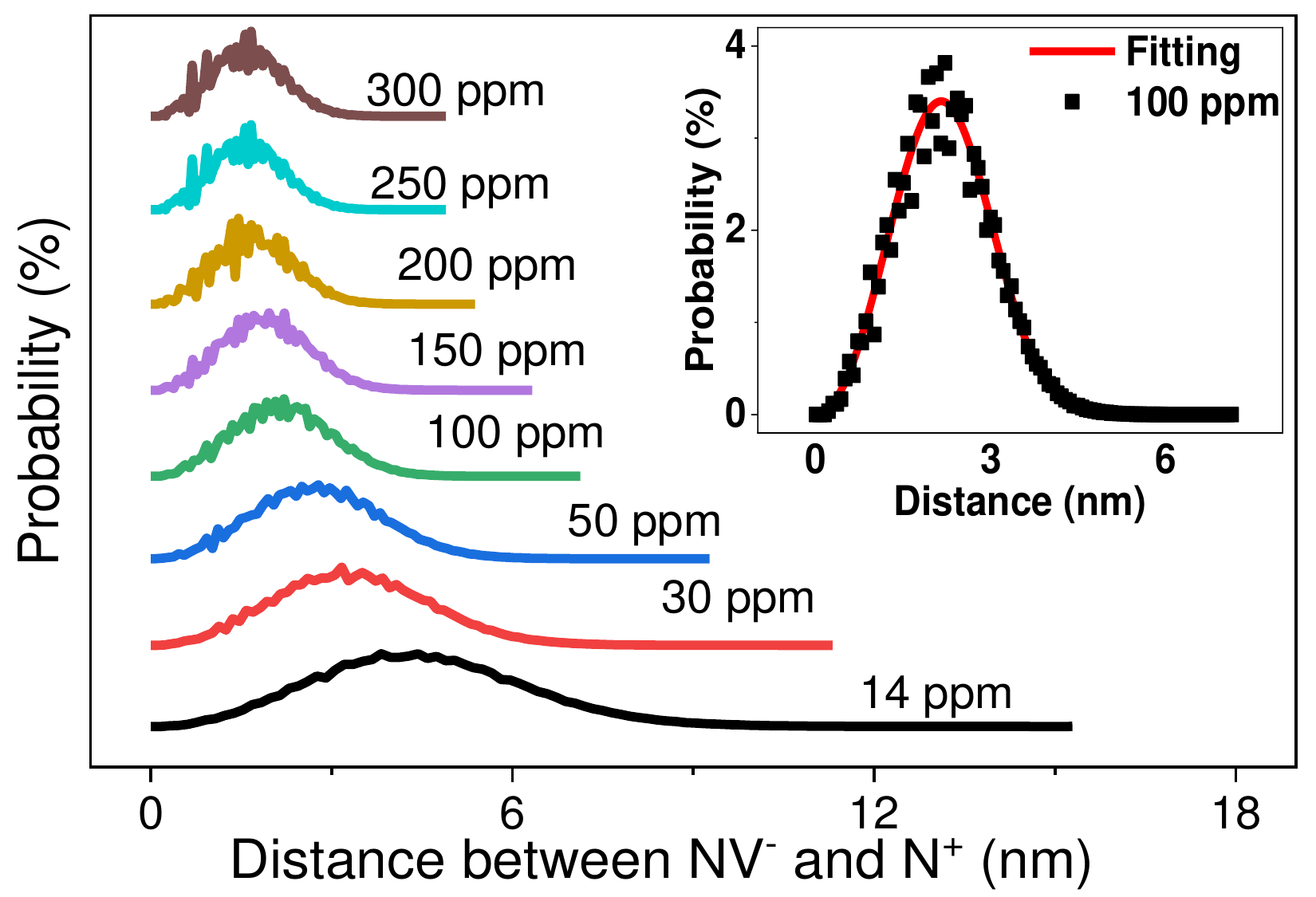}%
\caption{The probability distribution of the distance between the paired NV$ ^{-} $ and N$ ^{+} $ at different nitrogen concentrations. 
Inset: For [N] = 100 ppm, [NV$ ^{-} $] = 3 ppm, the probability distribution of 
the distance of NV$ ^{-} $-N$ ^{+} $ pair and fitting as Eq.\,\eqref{probability02}.}
\label{Fig_S2}
\end{figure}
The simulation results for samples with different nitrogen concentrations are shown in Fig. 3 of the main text. 
The probability distribution of the distance between NV$^{-}$-N$^{+}$ pairs are presented in Fig.\,\ref{Fig_S2}. 
Specifically, The inset Fig.\,\ref{Fig_S2} depicts the distribution probability of the distance between the NV$ ^{-} $-N$ ^{+} $ pairs in the sample with [N] = 100\, ppm and [NV$ ^{-} $] = 3\, ppm, while the red line represents the distribution probability fitted using Eq.\,\eqref{probability02}.
\pagebreak
\subsection{Laser power dependence and comparison with experimental ODMR spectra}
In order to compare the numerical simulation with the experimental ODMR spectra quantitatively, one needs to take into account the intrinsic spin dephasing  ($T_{2}^{*}$), the microwave power broadening (characterized by the Rabi frequency $\Omega_{R}$), as well as the light narrowing effect \cite{Budker_narrowing}. 
The spin dephasing process brings about a distribution of the zero-field splitting $D_{gs}$, with a characteristic width of  $1/T_{2}^{*}$ = 2 MHz in our experiments, then the corresponding Hamiltonian of the $i$-th NV$^{-}$ center becomes:
\begin{equation}\label{probabilitynew01}
\hat{H}_{i}=\left(D_{g s}+\sigma_{i} + \Pi_{z}^{i}\right) S_z^2+\Pi_{x}^{i}\left(S_y^2-S_x^2\right) +\Pi_{y}^{i}\left(S_x S_y+S_y S_x\right),
\end{equation}
where $\sigma_{i}$ obeys the N(0,$1/T_{2}^{*}$) Gaussian distribution.
By diagonalizing these Hamiltonians and counting the eigenvalues, we get the probability distribution of the resonant frequencies P($[N]$,$1/T_{2}^{*}$,$f_{MW}$):
\begin{equation}\label{probabilitynew02}
P([N],1/T_{2}^{*},f_{MW})=\sum_{f_{0}} A([N],1/T_{2}^{*},f_{0})\delta(f_{MW}-f_{0}),
\end{equation}
where $A([N], 1/T_{2}^{*}, f_{0})$ is the normalized amplitude, $f_{0}$ is the eigenvalues of Hamiltonians.

For the microwave power broadening effect, as a microwave pulse with $\Delta$ detuning has the probability of $\dfrac{\Omega_{R}^2}{\Omega_{R}^2+(2\pi\Delta)^2}$ to excite the NV spin transition, the Eq.\,\eqref{probabilitynew02} can be rewritten as:
\begin{equation}\label{probabilitynew03}
P([N],1/T_{2}^{*},\Omega_{R},f_{MW})=\sum_{f_{0}} A([N],1/T_{2}^{*},f_{0})\dfrac{\Omega_{R}^2}{\Omega_{R}^2+(f_{MW}-f_{0})^2},
\end{equation}

The following parameters are used to reproduce the ODMR spectra of the measured ND. NV$^{-}$ concentration [NV$^{-}$] = 3 ppm, nitrogen concentration [N] = 200 ppm, $1/T_{2}^{*}$ = 2 MHz, and Rabi frequency $\Omega_{R}$= 2$\pi\times$4 MHz. As shown in Fig.~4~(a-b) of the main text, the simulation results agree well with the experimental ODMR spectra.

Let us now consider the laser power dependence, which can be divided into two steps. 
First, we need to establish the relationship between the laser intensity and the ratio of the remaining NV$^{-}$ centers. To do this, we use the experimentally measured PL spectra, in which the ZPL of both the NV$^{-}$ (637 nm) and NV$^{0}$ (574 nm) centers can be easily quantified. 
Secondly, with the number of the remaining NV$^{-}$ centers, the above calculation can be performed again and ODMR spectra at the measured laser powers are obtained. The simulated spectra are then fitted with Eq.\,\eqref{probability16} and the key parameters of the ODMR spectra, splitting S($[N]$, $T_{2}^{*}$) and width W($[N]$, $T_{2}^{*}$), are obtained, as shown in Fig.~4~(c-d) of the main text.

By directly comparing the numerical and experimental results on laser power dependence, we find that the splitting values of both sides agree well. However, there is an apparent gap in ODMR width values. This discrepancy may be mainly caused by the laser narrowing effect. As shown in reference \cite{Budker_narrowing}, the width of microwave power-broadened ODMR spectra will be narrowed as laser excitation effectively decreases the NV spin relaxation time ($T_1$).
To account for the effect of light narrowing, we perform the above numerical simulation again but without the effect of microwave power broadening.
The obtained ODMR widths are treated as intrinsic widths (only the spin dephasing and NV charge related effects are considered). Next, we use equation 1 from the citation\,\cite{Budker_narrowing} to refine the laser power dependence of the ODMR width, as plotted in Fig.~4~(d) of the main text.

\subsection{Laser-induced charge conversion between NV$^{-}$and NV$^{0}$}
\begin{figure}[H]
\centering
\includegraphics[width=0.8\textwidth]{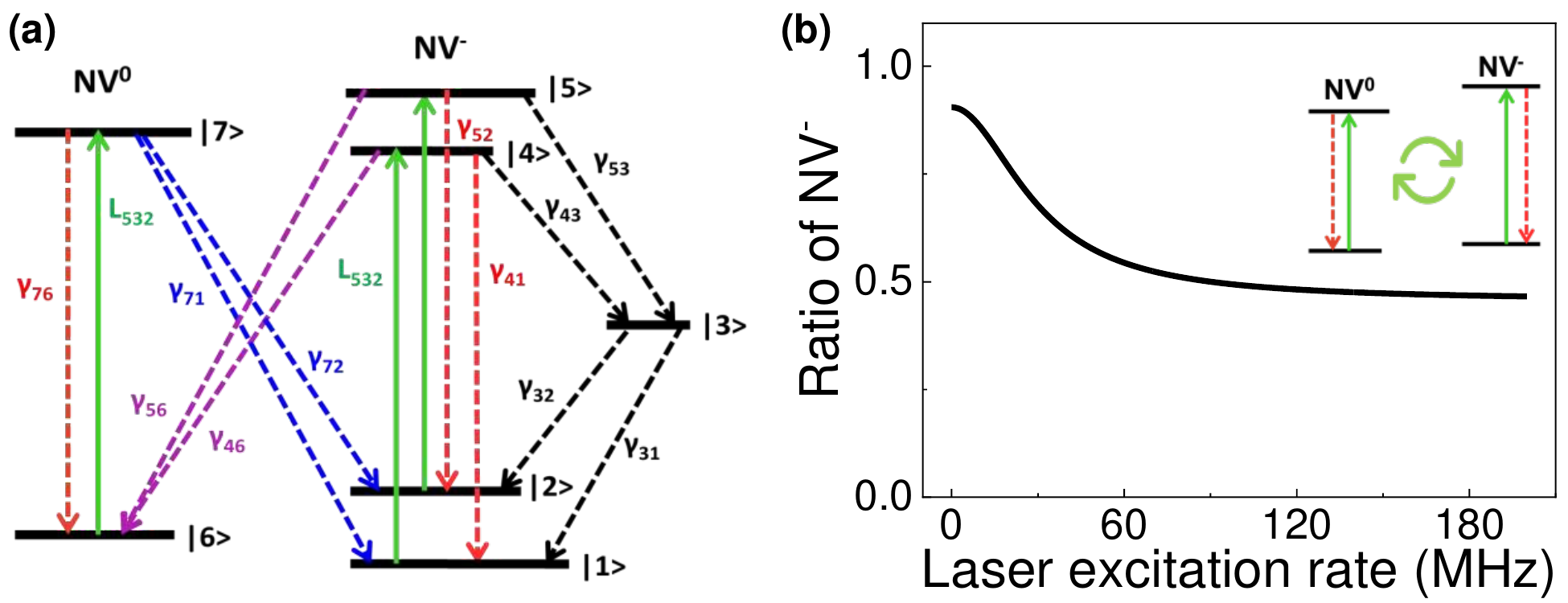}%
\caption{(a) The energy levels  of NV$ ^{-}$/NV$ ^{0}$ employed in our model of charge conversion.
(b) simulated remaining NV$^{-}$ center ratio at different laser excitation rates (different laser powers).}
\label{Fig_S3}
\end{figure}

The laser-induced charge conversion of NV centers has been extensively studied \cite{aslam2013photo,chen2017near,yuan2020charge,siyushev2013optically,savinov2022diamond,roberts2019spin,giri2019selective}. 
In our model, we use five levels to describe the NV$ ^{-} $ spin states, including two ground states ($\ket{1}$ and $\ket{2}$), two excited states($\ket{4}$ and $\ket{5}$), and an "effective" singlet state ($\ket{3}$). This model also includes the ground and excited states of NV$ ^{0} $ ($\ket{6}$ and $\ket{7}$). We use $\gamma_{ij}$ to represent the rate of spontaneous emission from energy level $\ket{i}$ to energy level $\ket{j}$, as shown in Fig.\,\ref{Fig_S3}~(a).
We define the $\Gamma_{ion} = \gamma_{56}+\gamma_{46}$, $\Gamma_{rec} = \gamma_{72}+\gamma_{71}$, and $\Gamma_{PL} = \gamma_{52}+\gamma_{51}+\gamma_{76}$ as the ionization, recombination and photoluminescence rates, respectively. The values of these parameters are listed in Table \,\ref{tab:lasertoPara}. 

\begin{table}[H]
\caption{Parameters used in the numerical simulation of laser power dependence, most of the values are taken from\,\cite{roberts2019spin,yuan2020charge}.}
\begin{ruledtabular}
\begin{tabular}{cccc}
\textrm{Parameter}    &\textrm{Rate (MHz)}      &\textrm{Parameter}     &\textrm{Rate (MHz)} \\
\colrule
$\gamma_{41}$         &80                       &$\gamma_{52}$          &80 \\     
$\gamma_{32}$         &25                       &$\gamma_{31}$          &75 \\   
$\gamma_{76}$         &20                       &$\gamma_{43}$          &15\\
$\gamma_{53}$         &45                       &$\gamma_{46}$          &20\\
$\gamma_{56}$         &20                       &$\gamma_{71}$          &5\\
$\gamma_{72}$         &10                       &                       &             \\
\end{tabular}
\end{ruledtabular}
\label{tab:lasertoPara}
\end{table}
We assume that the laser excitation rate is the same for NV$^-$ and NV$^0$.
This system can be described by the Lindblad master equation\,\cite{Lindblad1976}:
\begin{eqnarray}\label{masterequation}
\dot\rho=\frac{1}{i\hbar}[H,\rho]
+\sum_{<ij>}\gamma_{ij}\left(L_{ij}\rho L_{ij}^\dagger-\frac{1}{2}\left\{L_{ij}^\dagger L_{ij},\rho\right\}\right),
\end{eqnarray}
where $\rho$ is the density matrix of the system, $H = L_{532}(\ket{4}\bra{1}+\ket{7}\bra{6}+\ket{5}\bra{2}+ h.c.)$ is the Hamiltonian of the composite system with the laser excitation rate( $L_{532}$) in the interaction picture,
$L_{ij}=\ket{j}\bra{i}$ is a set of operators describing the dissipating part of the dynamics with the rate $\gamma_{ij}$. 
Using these equations, we perform numerical simulations\,\cite{tan1999computational} in different regions of the laser excitation rate and find the steady-state solution($\dot\rho=0$). The ratio of the remaining NV$^{-}$ center is equal to $\rho_{11}+\rho_{22}+\rho_{33}+\rho_{44}+\rho_{55}$. 
In the framework of the master equation theory, the laser excitation rate is proportional to the laser intensity. Therefore, the ratio decreases with the increase of the laser intensity (laser excitation rate). Fig.~\ref{Fig_S3} (b) shows the remaining NV$^{-}$ center ratio at different laser excitation rates (different laser powers).
\pagebreak
\section{ Extended data: ODMR of nanodiamonds and microdiamods}
\label{subsec:result}
\begin{figure}[H]
\centering
\includegraphics[width=0.9\textwidth]{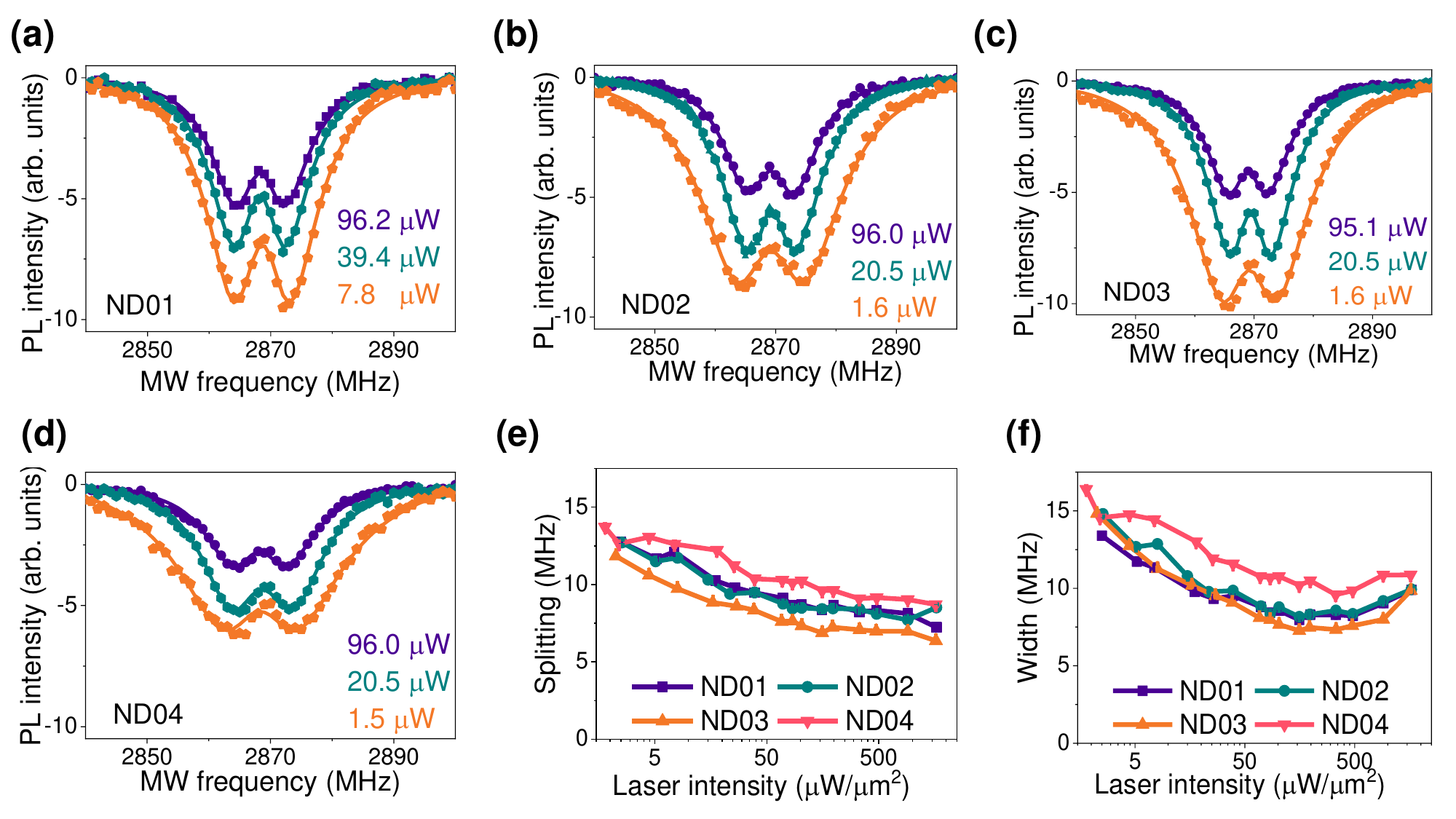}
\caption{The ODMR spectra of nanodiamonds under different laser powers.
(a-d) ODMR spectra of four nanodiamond particles.
(e) and (f) show the splitting and width of the measured ODMR spectra, respectively.}
\label{Fig_S4}
\end{figure}

\begin{figure}[H]
\centering
\includegraphics[width=0.9\textwidth]{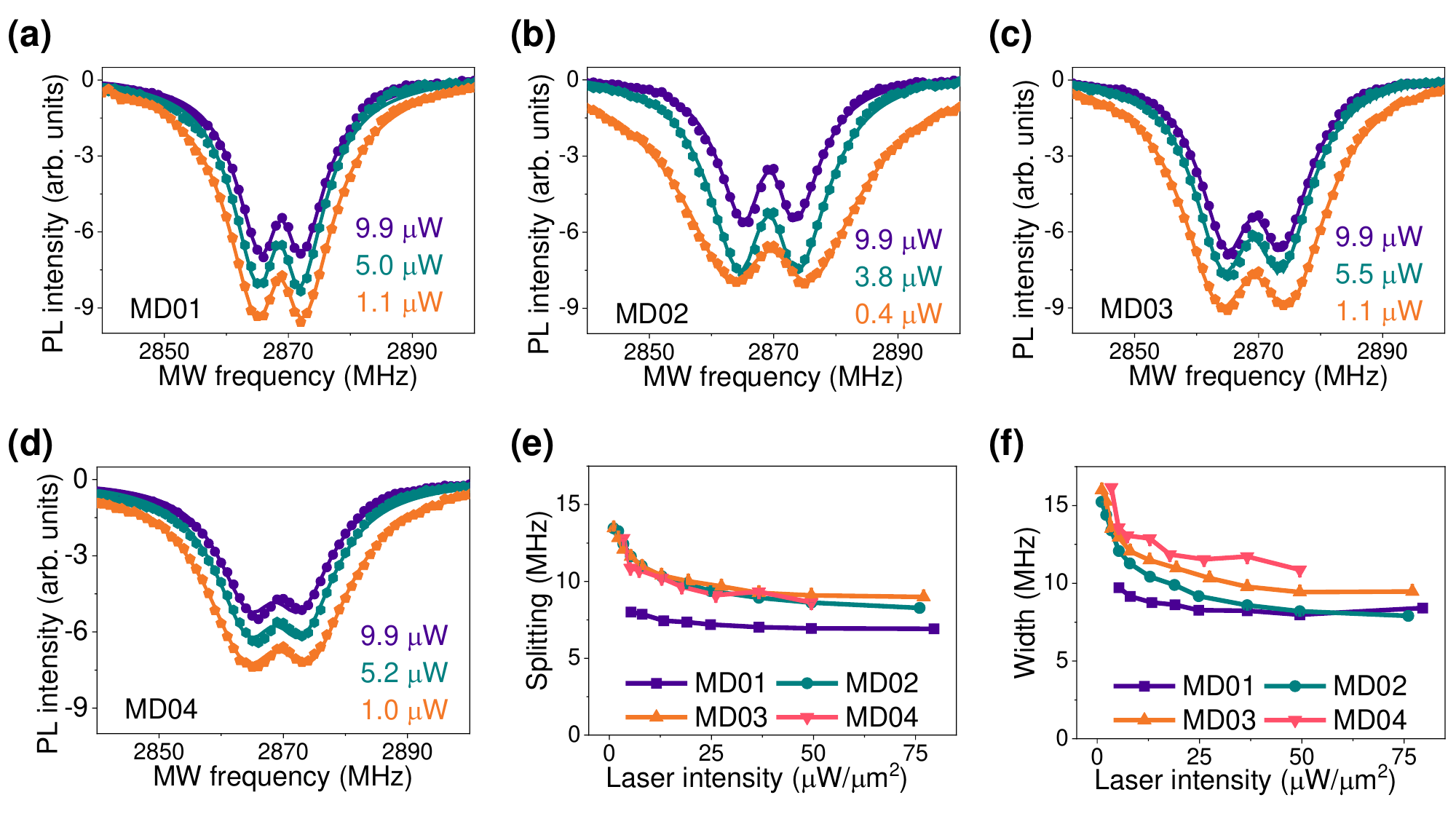}
\caption{The ODMR spectra of microdiamonds under different laser powers.
(a-d) ODMR spectra of four microdiamond particles.
(e) and (f) the splitting and width of the measured ODMR spectra.}
\label{Fig_S5}
\end{figure}

Figure\,\ref{Fig_S4} displays the ODMR spectra of four different nanodiamonds. 
In addition, the ODMR spectra of four microdiamonds are shown in Fig.\,\ref{Fig_S5}. 
The signals are normalized to the photon counts obtained at a fat detuned MW frequency. 
To estimate the splitting and width of these spectra, 
we use Eq.\,\eqref{probability16} from Section \,\ref{Theory} to fit the ODMR data.
The laser-induced narrowing and decreasing of splitting are observed for all measured nanodiamonds and microdiamonds.

\begin{figure}[H]
\centering
\includegraphics[width=0.8\textwidth]{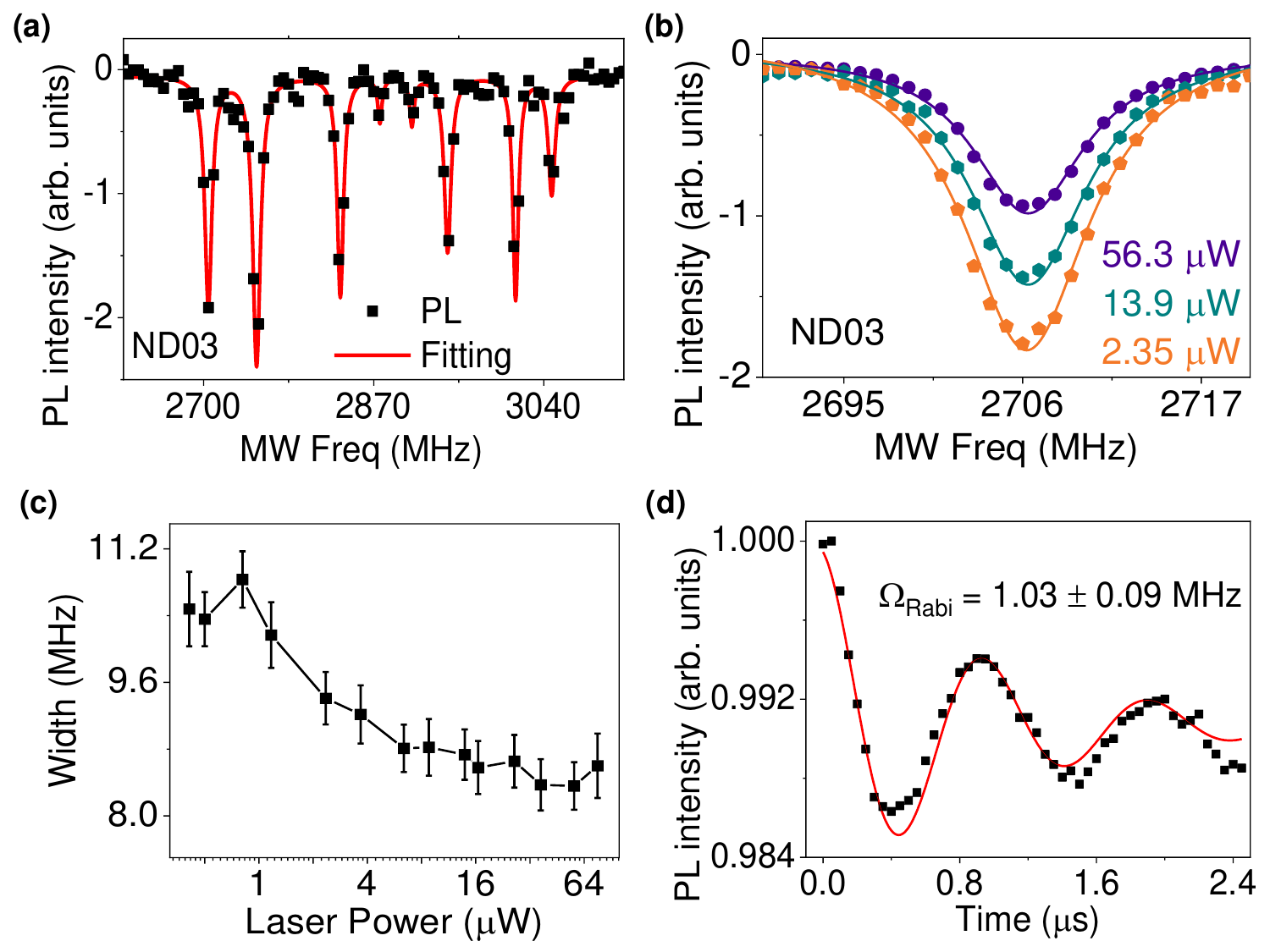}
\caption{ODMR spectra of an ND under an external magnetic field of about 58 gauss.
(a) Full ODMR spectra of the measured ND.
(b) The first resonant dip at different laser powers.
(c) Laser power dependence of the ODMR width.
(d) Rabi oscillation of NV spins, driving by resonant microwave pulses of the same power.}
\label{Fig_S6}
\end{figure}

The effect of the local electric field can be suppressed by an external magnetic field. 
As shown in Fig.~\ref{Fig_S6}~(a), the ODMR spectrum of an ND at an external magnetic field of about 58 gauss exhibits 8 resonant dips corresponding to the four possible NV orientations in the diamond crystal. The laser power dependence of the first resonant dip is measured and summarized in Fig.~\ref{Fig_S6}~(c). 
The width of the ODMR dip decreases slightly with increasing laser power, which is mainly due to the effects of light narrowing, as reported by Jensen \emph{et al}\,\cite{Budker_narrowing}.
To estimate the effective microwave power of the ODMR measurements, we also measured the Rabi oscillation of the same NV spins (Fig.~\ref{Fig_S6}~(d)).
\pagebreak
\pagebreak
\begin{table}[H]
\caption{Fitting parameters for the ODMR of nanodiamonds and microdiamonds.}
\begin{ruledtabular}
\begin{tabular}{cccccccc}
\textrm{Sample name}     &\textrm{Laser power ($\mathrm{\mu}$W)}     &\textrm{$C_{0}$}      
&\textrm{$C$} 
&\textrm{$n$}      &\textrm{$k$ (MHz$^{n-1}$)}
&\textrm{$\delta$ (MHz)}      &\textrm{$D_{gs}$ (MHz)}\\
\colrule
ND01&7.8&1.15e-3&-9.98&3.46&3.68e1&3.42&2868.64\\     
ND01&39.4&9.55e-4&-7.56&3.57&3.48e1&2.94&2868.28\\   
ND01&96.2&-2.38e-4&-5.43&3.64&3.40e1&3.01&2868.29\\
ND02&1.6&2.11e-3&-1.86e1&3.63&7.82e1&5.19&2869.22\\     
ND02&20.5&1.16e-3&-6.55&3.55&3.10e1&3.24&2869.34\\   
ND02&96.0&2.79e-4&-8.49&4.11&5.47e1&3.35&2869.20\\
ND03&1.6&9.93e-4&-5.79&3.15&2.41e1&4.16&2869.21\\     
ND03&20.5&7.35e-4&-3.17&3.20&1.48e1&2.61&2869.40\\   
ND03&95.1&1.71e-4&-1.88&3.27&1.39e1&2.69&2868.95\\
ND04&1.5&2.18e-3&-3.48&2.83&2.56e1&5.40&2868.39\\     
ND04&20.5&1.04e-3&-4.61e1&4.64&2.51e2&4.63&2868.68\\   
ND04&96.0&4.56e-6&-3.98&3.65&4.18e1&3.80&2868.28\\
MD01&1.1&9.73e-5&-1.82&2.60&8.83&3.02&2868.70\\   
MD01&5.0&5.92e-4&-1.70&2.74&8.74&2.57&2868.55\\   
MD01&9.9&5.80e-4&-1.39&2.77&8.41e1&2.51&2868.82\\
MD02&0.4&-2.65e-3&-6.29e1&2.95&3.53e1&5.49&2869.50\\     
MD02&3.8&2.45e-3&-5.99&3.17&2.77e1&3.31&2869.19\\   
MD02&9.9&1.35e-3&-3.63&3.17&2.14e1&2.58&2869.29\\
MD03&1.1&2.56e-3&-3.52e1&3.98&1.32e2&5.46&2869.32\\     
MD03&5.5&2.78e-3&-2.28e1&4.12&9.40e1&4.25&2869.03\\   
MD03&9.9&1.95e-3&-1.54e1&4.04&7.12e1&3.79&2869.32\\
MD04&1.0&-1.64e-3&-7.53&3.44&4.52e1&5.61&2869.10\\     
MD04&5.2&-6.14e-4&-3.65&3.35&2.57e1&4.61&2868.73\\   
MD04&9.9&-6.50e-4&-3.31&3.48&2.68e1&4.38&2868.65\\
\end{tabular}
\end{ruledtabular}
\label{tab:laserPara}
\end{table}

\bibliographystyle{apsrev4-2-2}
\bibliography{literature.bib}